\documentclass[twocolumn,floatfix]{revtex4}
\pdfoutput=1
\usepackage{graphicx}
\usepackage{dcolumn}
\usepackage{longtable}
\usepackage{amsmath}
\usepackage{amssymb}

\begin{document}

\title
{Analytic predictions for nuclear shapes, the prolate dominance and the prolate-oblate shape transition
in the proxy-SU(3) model}

\author
{Dennis Bonatsos$^1$, I. E. Assimakis$^1$, N. Minkov$^2$, Andriana Martinou$^1$, S. Sarantopoulou$^1$,
R. B. Cakirli$^3$, R. F. Casten$^{4,5}$, K. Blaum$^6$}

\affiliation
{$^1$Institute of Nuclear and Particle Physics, National Centre for Scientific Research 
``Demokritos'', GR-15310 Aghia Paraskevi, Attiki, Greece}

\affiliation
{$^2$Institute of Nuclear Research and Nuclear Energy, Bulgarian Academy of Sciences, 72 Tzarigrad Road, 1784 Sofia, Bulgaria}

\affiliation
{$^3$ Department of Physics, University of Istanbul, Istanbul, Turkey} 

\affiliation 
{$^4$ Wright Laboratory, Yale University, New Haven, Connecticut 06520, USA}

\affiliation
{$^5$ Facility for Rare Isotope Beams, 640 South Shaw Lane, Michigan State University, East Lansing, MI 48824 USA}

\affiliation
{$^6$ Max-Planck-Institut f\"{u}r Kernphysik, Saupfercheckweg 1, D-69117 Heidelberg, Germany}

\begin{abstract}

Using a new approximate analytic parameter-free proxy-SU(3) scheme, we make simple predictions of shape observables for deformed nuclei, namely $\gamma$ and $\beta$ deformation variables, the global feature of prolate dominance and the locus of the prolate-oblate shape transition. The predictions are compared with empirical results.

\end{abstract}

\maketitle

\section{Introduction}

The existence of both prolate (cigar shaped) and oblate (pancake shaped) deformed nuclei,
the possible transitions between the two shapes, as well as the experimentally observed dominance 
of prolate over oblate shapes in the ground state bands of even nuclei has been a focus of attention 
for decades and from many different viewpoints.

It is the purpose of this paper to exploit a new approximate SU(3) symmetry to obtain analytic, parameter-free predictions, essentially by inspection,  of the dominance of prolate shapes in atomic nuclei and of the locus of the prolate-oblate transition.  The new symmetry scheme, called a proxy-SU(3), is similar in spirit to pseudo-SU(3) \cite{pseudo1,pseudo2,Ginocchio} but involves a different ansatz in order to obtain a valence space symmetry. 

The proxy-SU(3) concept has been introduced and vetted in Ref. \cite{proxy} where it was shown that the Nilsson diagrams for well-deformed nuclei obtained with the new symmetry are very similar to the traditional Nilsson diagrams.
Briefly, proxy-SU(3) simply replaces all but one of the intruder unique parity orbitals in medium and heavy mass nuclei by the highest $j$-orbit from the next lower shell that is very similar in spatial overlap and which has identical angular momentum projection properties.  
This produces a new proxy set of orbits, very similar to the original set, that constitutes a full oscillator shell (all even or all odd orbital angular momenta up to some maximum value, such as the s,d, and g orbits, with $l = 0$, 2, and 4 and with both $j = l +1/2$ and $j = l - 1/2$ total angular momenta). Such a set of orbits has symmetry U(X) where X is  half the number of nucleons in the proxy shell (30 nucleons for 50-82 and 42 for 82-126). Bear in mind that the highest lying unique parity orbital, like 11/2[505] in the 50-82 shell, which contains two particles, is lost in the proxy shell approximation. 

Having such a symmetry allows a number of simple predictions resulting solely from the group structure of the symmetry itself. The motivation for this approximate symmetry and its detailed character are further described in \cite{proxy} and summarized at the beginning of Section II below. These predictions arise simply from filling the nucleon orbitals in a deformed quadrupole field, and the consequent changes in ground state irreducible representations (irreps) for the relevant group (see below).

Over the years, there have been many efforts to understand nuclear shapes and the locus of prolate and oblate shapes in nuclei from many different perspectives. 
Microscopic calculations have evolved from early applications of the pairing plus quadrupole model 
to the prolate-oblate difference \cite{Kumar1} and the prolate-oblate transition \cite{Kumar2} to recent 
self-consistent Skyrme Hartree-Fock plus BCS calculations \cite{Sarriguren} and Hartree-Fock-Bogoliubov calculations 
\cite{Robledo,Nomura83,Nomura84} studying the structural evolution in neutron-rich Yb, Hf, W, Os, and Pt isotopes, 
reaching the conclusion that $N \approx 116$ nuclei in this region can be identified as the transition point between prolate and 
oblate shapes. In a related projected shell model study \cite{Sun}, a rotation-driven prolate-to-oblate shape phase transition has
been found in $^{190}$W.  
The prolate-oblate shape phase transition has been considered \cite{Thiamova,Bettermann,Zhang40}
within the O(6) symmetry of the interacting boson model \cite{IA}. In particular, the O(6) symmetry has been considered \cite{Jolie1,Jolie2}  
as a critical point of the prolate-to-oblate shape phase transition within the interacting boson model. 
An analytically solvable prolate-to-oblate shape phase transition has been found \cite{Zhang85} within the SU(3) limit of the interacting 
boson model. The collection of data of the chain of even nuclei (differing by two protons or two neutrons) $^{180}$Hf, $^{182-186}$W, 
$^{188,190}$Os, $^{192-198}$Pt, considered in Ref. \cite{Zhang85}, suggests that the transition occurs between $^{190}$Os and
 $^{192}$Pt, in agreement with their theoretical predictions.  
The dominance of prolate over oblate nuclear shapes in the ground state bands of deformed even-even nuclei has been considered both in the framework of the Nilsson model \cite{Casten,Hamamoto}, as well as by studying the effects of the spin-orbit potential within the framework 
of the Nilsson-Strutinsky method \cite{Tajima64,Tajima702,Tajima86}. 
Nevertheless, the almost complete dominance of prolate over oblate deformations in the ground states of even-even nuclei is still considered as not adequately understood \cite{HM}. 

From the experimental point of view, $^{192}$Os \cite{Namenson} and $^{190}$W \cite{Alkhomashi} have been suggested as lying at the prolate-oblate border, with $^{194}$Os \cite{Wheldon} and $^{198}$Os \cite{Podolyak} having an oblate character. Data on nuclei from Hf to Pt, discussed in Ref. \cite{Linnemann}, also suggests that the transition occurs between $^{192}$Os and $^{194}$Pt.

In the present work, we consider nuclear shapes in terms of the standard variables $\gamma$ and $\beta$, as well as the prolate-oblate competition within the framework of the recently proposed \cite{proxy} parameter-free proxy-SU(3) symmetry in nuclei. Our main results are:

1) predictions of nuclear quadrupole deformations and axial asymmetry for deformed nuclei and a comparison with empirical results,

2) the dominance of prolate-over-oblate deformation,

3) the occurrence of the prolate-oblate transition at $N \approx 116$ in agreement with the data
in the W and Os chains of isotopes, while predictions are made for $Z<74$ (i.e., below W),  

4) predictions are made concerning the prolate-oblate transition in the region of the (yet unknown) neutron-deficient rare earths around $N \approx 72$. 

\section{The proxy-SU(3) scheme} 

A proxy-SU(3) symmetry scheme, applicable in heavy deformed nuclei, has been recently introduced \cite{proxy},
based on the asymptotic Nilsson wave functions $| N_q n_z \Lambda \Sigma \rangle$ \cite{Nilsson1,Nilsson2}, where $N_q$ is the total number of oscillator quanta
(the subscript $q$ is added in order to distinguish this number from the neutron number $N$), 
$n_z$ is the number of the oscillator quanta along the $z$-axis, $\Lambda$ is the $z$-projection of the orbital angular momentum, and $\Sigma$ 
is the $z$-projection of the spin. Nilsson orbitals in even-even nuclei are then denoted by $K[N_q n_z \Lambda]$, where $K$ is the projection of the total angular momentum on the $z$-axis, given by $K= \Lambda +\Sigma$. 

The key to the new scheme is the great similarity between Nilsson orbitals  differing by $\Delta K [\Delta N_q \Delta n_z \Delta \Lambda]=0[110]$. 
Proton-neutron 0[110] pairs were found to play a key role in the deformation of heavy nuclei,
especially in those with equal numbers of  valence protons and valence neutrons \cite{Cakirli,Karampagia}. It was subsequently realized that 0[110] orbitals can be used 
for the construction of a proxy-SU(3) scheme for heavy deformed nuclei, similar to the Elliott SU(3) symmetry \cite{Elliott1,Elliott2,Elliott3} appearing in light nuclei. In both cases, the standard Elliott notation $(\lambda, \mu)$ is used for the irreducible representations (irreps) of SU(3). 

The proxy-SU(3) scheme results in a description of nuclei in terms of SU(3) representations from a direct product of the proton and neutron spaces. For a given nucleus ($\lambda$, $\mu$) values are directly related to the number of valence nucleons and, for the highest weight state (the ground state), tend to grow with those numbers up to the middle of the shell.
The Elliott labels $\lambda$ and $\mu$ are known \cite{Evans,Castanos,Park} to be  connected to the shape variables of the collective model \cite{BM}.  This connection is achieved by 
employing a linear mapping between the eigenvalues of invariant operators of the two theories, 
namely between the invariants $\beta^2$ and $\beta^3\cos 3\gamma$ of the collective model
(where $\beta$ and $\gamma$ stand for the usual collective variables) and the invariants of SU(3), 
which are the second and third order Casimir operators of SU(3), respectively \cite{IA} (see Appendix A for further discussion). 
The mapping results in the angle collective variable $\gamma$ given by \cite{Castanos,Park}
\begin{equation}\label{g1}
\gamma = \arctan \left( {\sqrt{3} (\mu+1) \over 2\lambda+\mu+3}  \right),
\end{equation}
and in the square of the deformation parameter $\beta$ being proportional to the second order Casimir operator of SU(3) \cite{IA}, 
 \begin{equation}\label{C2} 
 C_2(\lambda,\mu)= {2 \over 3} (\lambda^2+\lambda \mu + \mu^2+ 3\lambda +3 \mu), 
\end{equation}
and given by \cite{Castanos,Park}
\begin{equation}\label{b1}
	\beta^2= {4\pi \over 5} {1\over (A \bar{r^2})^2} (\lambda^2+\lambda \mu + \mu^2+ 3\lambda +3 \mu +3), 
\end{equation}
where $A$ is the mass number of the nucleus and $\bar{r^2}$ is related to the dimensionless mean square radius \cite{Ring}, $\sqrt{\bar{r^2}}= r_0 A^{1/6}$. The dimensionless mean square radius 
is obtained by dividing the mean square radius, which is proportional to $A^{1/3}$, by the oscillator length, which grows as $A^{1/6}$ \cite{Ring}.
The constant $r_0$ is determined from a fit over a wide range of nuclei \cite{DeVries,Stone}. We use the value in Ref. \cite{Castanos}, $r_0=0.87$, in agreement to Ref. \cite{Stone}. 

Alternatively, one can use the invariants as formulated in Ref. \cite{Evans}. In that case, the expression resulting for $\beta^2$ is identical to Eq.~(\ref{b1}) (the only difference being that 
the last term in the paranthesis, $+3$, is missing), while for $\gamma$ the result reads 
\begin{equation}\label{g2}
\cos 3\gamma= {(\lambda-\mu)(\lambda+2\mu+3)(2\lambda+\mu+3) \over 2 (\lambda^2+ \mu^2+\lambda \mu 
+ 3\lambda + 3\mu)^{3/2}},
\end{equation}
in agreement to the result obtained in Ref. \cite{Troltenier}. 
It can be seen that Eqs.~(\ref{g1}) and~(\ref{g2}) yield almost identical 
results, except for values very close (less than one degree away) to 0 or to $\pi/3$, where 
Eq.~(\ref{g1}) still works without any problem, while Eq.~(\ref{g2}) fails, the reason being that 
the approximations involved in both cases induce small errors, which are insignificant if the tangent 
is used, but lead to violation of the condition $|\cos 3\gamma|\leq 1$ if the cosine is used. 

\begin{table*}[htb]

\caption{Highest weight SU(3) irreps (which are always unique) for U(n), n=6, 10, 15, 21
given in the columns labelled by hw, contained in the relevant U(n) irrep  for M valence protons or M valence neutrons,
compared to SU(3) irreps with the highest eigenvalue of the second order Casimir operator of SU(3),
given in the columns labelled by C. Above the U(n) algebra, the relevant shell of the shell model 
and the corresponding proxy-SU(3) shell are given.   
The upper half of C columns is identical to that of the corresponding hw column. The lower half of the C columns is a mirror image of their upper half, while in the lower half of hw columns violations of the mirror symmetry appear, indicated by boldface characters. The code UNTOU3 \cite{code} has been used for producing these results. Note that the proxy-SU(3) scheme omits the highest $K$ Nilsson orbital from the unique parity orbit (e.g., 13/2[606] for the 82-126 shell) and therefore the sizes of the proxy sdg and proxy pfh shells are 30 and 42 nucleons instead of the normal 32 and 44 nucleons for the 50-82 and 82-126 shells, respectively. Exactly at mid-shell (n particles in the case of U(n)), there exist two irreps possessing the same maximum eigenvalue of the Casimir operator, the highest weight irrep and its mirror image. For exampe, in U(15) for n=15 the highest weight leads to the (19,7) irrep, while the highest eigenvalue of the Casimir operator is possessed by the (19,7) and (7,19) irreps. 
See Section \ref{p-h} and Appendix B for further discussion.
}

\bigskip
\begin{tabular}{ r l r r r r r r r r  }

\hline

\hline
   &             & 8-20 & 8-20 & 28-50 & 28-50& 50-82 & 50-82 & 82-126& 82-126\\
   &             & sd   &  sd  & pf    &  pf  & sdg   &  sdg  &  pfh  & pfh   \\
M  & irrep       & U(6) & U(6) & U(10) & U(10)& U(15) & U(15) & U(21) & U(21) \\
   &             & hw   & C    & hw    &  C   & hw    &  C    & hw    &   C   \\
 0 &             &(0,0) &(0,0) &(0,0)  &(0,0) &(0,0)  &(0,0)  &(0,0)  &(0,0)  \\  
 1 & [1]         &(2,0) &(2,0) & (3,0) &(3,0) & (4,0) &(4,0)  & (5,0) &(5,0)  \\
 2 & [2]         &(4,0) &(4,0) & (6,0) &(6,0) & (8,0) &(8,0)  &(10,0) &(10,0) \\
 3 & [21]        &(4,1) &(4,1) & (7,1) &(7,1) &(10,1) &(10,1) &(13,1) &(13,1) \\
 4 & [$2^2$]     &(4,2) &(4,2) & (8,2) &(8,2) &(12,2) &(12,2) &(16,2) &(16,2) \\
 5 & [$2^2$1]    &(5,1) &(5,1) &(10,1) &(10,1)&(15,1) &(15,1) &(20,1) &(20,1) \\
 6 & [$2^3$]     &(6,0) & (0,6)&(12,0) &(12,0)&(18,0) &(18,0) &(24,0) &(24,0) \\
 7 & [$2^3$1]  &{\bf(4,2)}&(1,5)&(11,2)&(11,2)&(18,2) &(18,2) &(25,2) &(25,2) \\
 8 & [$2^4$]     &(2,4) & (2,4)&(10,4) &(10,4)&(18,4) &(18,4) &(26,4) &(26,4) \\
 9 & [$2^4$1]    &(1,4) & (1,4)&(10,4) &(10,4)&(19,4) &(19,4) &(28,4) &(28,4) \\
10 & [$2^5$]     &(0,4) & (0,4)&(10,4) &(4,10)&(20,4) &(20,4) &(30,4) &(30,4) \\
11 & [$2^5$1]   &(0,2)&(0,2)&{\bf(11,2)}&(4,10)&(22,2)&(22,2) &(33,2) &(33,2) \\
12 & [$2^6$]    &(0,0)&(0,0)&{\bf(12,0)}&(4,10)&(24,0)&(24,0) &(36,0) &(36,0) \\
13 & [$2^6$1]   &     &     &{\bf(9,3)} &(2,11)&(22,3)&(22,3) &(35,3) &(35,3) \\
14 & [$2^7$]    &     &     &{\bf(6,6)} &(0,12)&(20,6)&(20,6) &(34,6) &(34,6) \\
15 & [$2^7$1]   &     &     &{\bf(4,7)} &(1,10)&(19,7)& (7,19)&(34,7) &(34,7) \\
16 & [$2^8$]    &  &      & (2,8) & (2,8)&{\bf(18,8)} & (6,20)&(34,8) &(34,8) \\
17 & [$2^8$1]   &  &      & (1,7) & (1,7)&{\bf(18,7)} & (3,22)&(35,7) &(35,7) \\
18 & [$2^9$]    &  &      & (0,6) & (0,6)&{\bf(18,6)} & (0,24)&(36,6) &(36,6) \\
19 & [$2^9$1]   &  &      & (0,3) & (0,3)&{\bf(19,3)} & (2,22)&(38,3) &(38,3) \\
20 & [$2^{10}$] &  &      & (0,0) & (0,0)&{\bf(20,0)} & (4,20)&(40,0) &(40,0) \\
21 & [$2^{10}$1]&  &      &       &      &{\bf(16,4)} & (4,19)&(37,4) & (4,37) \\
22 & [$2^{11}$] &  &      &       & &{\bf(12,8)} & (4,18)&{\bf(34,8)} &(0,40)\\
23 & [$2^{11}$1]&  &      &       & &{\bf(9,10)} & (2,18)&{\bf(32,10)}&(3,38)\\
24 & [$2^{12}$] &  &      &       & &{\bf(6,12)} & (0,18)&{\bf(30,12)}&(6,36)\\
25 & [$2^{12}$1]&  &      &       & &{\bf(4,12)} & (1,15)&{\bf(29,12)}&(7,35)\\
26 & [$2^{13}$] &      &      &   &      &(2,12) & (2,12)&{\bf(28,12)}&(8,34)\\
27 & [$2^{13}$1]&      &      &   &      &(1,10) & (1,10)&{\bf(28,10)}&(7,34)\\
28 & [$2^{14}$] &      &      &   &      & (0,8) & (0,8) &{\bf(28,8)} &(6,34)\\
29 & [$2^{14}$1]&      &      &   &      & (0,4) & (0,4) &{\bf(29,4)} &(3,35)\\
30 & [$2^{15}$] &      &      &   &      & (0,0) & (0,0) &{\bf(30,0)} &(0,36)\\
31 & [$2^{15}$1]&      &      &   &      &       &       &{\bf(25,5)} &(2,33)\\
32 & [$2^{16}$] &      &      &   &      &       &       &{\bf(20,10)}&(4,30)\\
33 & [$2^{16}$1]&      &      &   &      &       &       &{\bf(16,13)}&(4,28)\\
34 & [$2^{17}$] &      &      &   &      &       &       &{\bf(12,16)}&(4,26)\\
35 & [$2^{17}$1]&      &      &   &      &       &       &{\bf(9,17)} &(2,25)\\ 
36 & [$2^{18}$] &      &      &   &      &       &       &{\bf(6,18)} &(0,24)\\
37 & [$2^{18}$1]&      &      &   &      &       &       &{\bf(4,17)} &(1,20)\\
38 & [$2^{19}$] &      &      &        &      &       &       &(2,16) &(2,16)\\
39 & [$2^{19}$1]&      &      &        &      &       &       &(1,13) &(1,13)\\
40 & [$2^{20}$] &      &      &        &      &       &       &(0,10) &(0,10)\\
41 & [$2^{20}$1]&      &      &        &      &       &       &(0,5)  &(0,5) \\
42 & [$2^{21}$] &      &      &        &      &       &       &(0,0)  &(0,0) \\
\hline

\end{tabular}

\end{table*} 

For a given nucleus, ($\lambda$, $\mu$) for the ground state are given through the outer product of the relevant proton and neutron SU(3) irreps [that is, by the sum of the proton and neutron 
$(\lambda, \mu)$s,  see below] and thus can be used to determine predicted values of both $\beta$ and $\gamma$.  One sees from these equations that:

1) within a given mass region (roughly constant $A$ value), $\beta$ increases with $\lambda$ and $\mu$ and therefore tends to maximize near mid-shell in agreement with the data;

2) for $\lambda \gg \mu$ one has $\gamma\approx 0$, while for $\lambda > \mu$ one has $30^{\rm{o}} > \gamma > 0$, 
corresponding to prolate shapes;

3) for $\lambda \ll \mu$ one has $\gamma\approx \arctan\sqrt{3} = 60^{\rm{o}}$, while for $\lambda < \mu$ one has $60^{\rm{o}} > \gamma > 30^{\rm{o}}$, corresponding to oblate shapes;

4) in the special case of $\lambda=\mu$ one has $\gamma=\arctan(1/\sqrt{3}) = 30^{\rm{o}}$, 
corresponding to maximal triaxiality. 

Below we will see how these ideas are borne out in practice and compare the predictions made 
with eqs.~(1) and~(3) with empirical results

The asymmetry of the dependence on $\lambda$, $\mu$ in Eq.~(1), the asymmetry in the $(\lambda,\mu)$ values themselves about mid-shell (see below) and the $A$ dependence in Eq.~(3), have an important consequence: the proxy-SU(3) predictions for $\beta$ and $\gamma$ and for prolate and oblate character are not symmetric about mid-shell, in contrast to some other models, leading to predictions that are in fact reflected in the data.

In the proxy-SU(3) scheme \cite{proxy}, the protons of the 50-82 shell live in a proxy sdg shell, having an approximate U(15) symmetry,
which is obtained by leaving out the (very high-lying) 11/2[505] orbital and replacing the rest of the 1h$_{11/2}$ subshell orbitals (1/2[550], 3/2[541], 5/2[532], 7/2[523], 9/2[514]) by their 0[110] counterparts \cite{Cakirli,Karampagia} 
(1/2[440], 3/2[431], 5/2[422],7/2[413], 9/2[404]), which form a 1g$_{9/2}$ subshell. 

Similarly, in the same scheme \cite{proxy}, the neutrons of the 82-126 shell live in a proxy pfh shell, having an approximate U(21) symmetry,
which is obtained by leaving out the (very high-lying) 13/2[606] orbital and replacing the rest of the 1i$_{13/2}$ subshell orbitals (1/2[660], 3/2[651], 5/2[642], 7/2[633], 9/2[624],11/2[615]) by their 0[110] counterparts \cite{Cakirli,Karampagia}
(1/2[550], 3/2[541], 5/2[532],7/2[523], 9/2[514], 11/2[505]), which form a 1h$_{11/2}$ subshell.

\begingroup
\begin{table*}[htb]

\caption{Most leading SU(3) irreps \cite{Elliott1,Elliott2} for nuclei with protons in the 50-82 shell 
and neutrons in the 82-126 shell. Boldface numbers indicate nuclei with $R_{4/2}=E(4^+_1)/E(2_1^+) \geq 2.8$, 
while * denotes nuclei with $2.8 > R_{4/2} \geq 2.5$, and ** labels a few nuclei with  $R_{4/2}$ ratios slightly below 2.5, 
shown for comparison, while no irreps are shown for any other nuclei with $R_{4/2}<2.5$. For the rest of the nuclei shown 
(using normal fonts and without any special signs attached) the $R_{4/2}$ ratios are still unknown \cite{ENSDF}. Irreps corresponding to oblate shapes are underlined. 
}

\bigskip

\begin{tabular}{ r r r | r r r r r r r r r r r r }

  &  &              & Ba    &  Ce    &  Nd    &  Sm    &   Gd   &  Dy    &  Er    &  Yb    &  Hf    &   W    &   Os   &  Pt     \\
  &  &  $Z$         &  56   & 58     & 60     & 62     & 64     & 66     & 68     & 70     & 72     & 74     & 76     & 78      \\
  &  &$Z_{val}$     &   6   &  8     &  10    &  12    &  14    &  16    &  18    &  20    & 22     &  24    &   26   &   28    \\
$N$ &$N_{val}$& irrep        &(18,0) & (18,4) &  (20,4)& (24,0) & (20,6) & (18,8) & (18,6) & (20,0) & (12,8) & (6,12) & (2,12) & (0,8)   \\   

\hline

88 & 6  &(24,0) &(42,0)*&(42,4)* &(44,4)* &        &        &        &        &        &        &        &        &         \\
90 & 8  &(26,4) &{\bf(44,4)}&{\bf(44,8)}&\bf{(46,8)}&\bf{(50,4)}&\bf{(46,10)}&\bf{(44,12)}&(44,10)*&(46,4)* &(38,12)*&  &   &     \\
92 & 10 &(30,4) &{\bf (48,4)}&\bf{(48,8)}&\bf{(50,8)}&\bf{(54,4)}&\bf{(50,10)}&\bf{(48,12)}&\bf{(48,10)}&\bf{(50,4)}&(42,12)* &  &  &  \\
94 & 12 &(36,0) &(54,0) &{\bf (54,4)}&\bf{(56,4)}&\bf{(60,0)}&\bf{(56,6)}&\bf{(54,8)}&\bf{(54,6)}&\bf{(56,0)}&\bf{(48,8)}&\bf{(42,12)}&(38,12)*&  \\
96 & 14 &(34,6) &(52,6) &(52,10) &{\bf (54,10)}&\bf{(58,6)}&\bf{(54,12)}&\bf{(52,14)}&\bf{(52,12)}&\bf{(54,6)}&\bf{(46,14)}&\bf{(40,18)}&(36,18)*& \\
98 & 16 &(34,8) &(52,8) &(52,12) &(54,12) &\bf{(58,8)}&\bf{(54,14)}&\bf{(52,16)}&\bf{(52,14)}&\bf{(54,8)}&\bf{(46,16)}&\bf{(40,20)}&(36,20)*&   \\
100& 18 &(36,6) &(54,6) &(54,10) &(56,10) &(60,6)  &\bf{(56,12)}&\bf{(54,14)}&\bf{(54,12)}&\bf{(56,6)}&\bf{(48,14)}&\bf{(42,18)}&\bf{(38,18)}&(36,14)* \\
102& 20 &(40,0) &(58,0) &(58,4)  &(60,4)  &(64,0)  &\bf{(60,6)}&\bf{(58,8)}&\bf{(58,6)}&\bf{(60,0)}&\bf{(52,8)}&\bf{(46,12)}&\bf{(42,12)}&(40,8)*  \\
104& 22 &(34,8) &(52,8) &(52,12) &(54,12) &(58,8)  &(54,14) &\bf{(52,16)}&\bf{(52,14)}&\bf{(54,8)}&\bf{(46,16)}&\bf{(40,20)}&\bf{(36,20)} &(34,16)* \\
106& 24 &(30,12)&(48,12)&(48,16) &(50,16) &(54,12) &(50,18) &(48,20) &\bf{(48,18)}&\bf{(50,12)}&\bf{(42,20)}&\bf{(36,24)}&\bf{(32,24)}&(30,20)* \\
108& 26 &(28,12)&(46,12)&(46,16) &(48,16) &(52,12) &(48,18) &(46,20) &(46,18) &\bf{(48,12)}&\bf{(40,20)}&\bf{(34,24)}&\bf{(30,24)}&(28,20)* \\
110& 28 &(28,8) &(46,8) &(46,12) &(48,12) &(52,8)  &(48,14) &(46,16) &(46,14) &(48,8)  &\bf{(40,16)}&\bf{(34,20)}&\bf{(30,20)}&(28,16)* \\
112& 30 &(30,0) &(48,0) &(48,4)  &(50,4)  &(54,0)  &(50,6)  &(48,8)  &(48,6)  &(50,0)  &\bf{(42,8)}&\bf{(36,12)}&\bf{(32,12)}&(30,8)** \\
114& 32 &(20,10)&(38,10)&(38,14) &(40,14) &(44,10) &(40,16) &(38,18) &(38,16) &(40,10) &(32,18) &\bf{(26,22)}&\bf{(22,22)}&(20,18)**\\
116& 34 &(12,16)&(30,6) &(30,10) &(32,10) &(36,6)  &(32,12) &(30,14) &(30,12) &(32,6)  &(24,14) &$\underline{(18,28)*}$&  $\underline{\bf{(14,28)}}$&$\underline{(12,24)**}$\\
118& 36 &(6,18) &(24,18)&(24,22) &(26,22) &(30,18) &(26,24) &(24,16) &(24,24) &(26,18) &$\underline{(18,26)}$ &$\underline{(12,30)}$ &$\underline{(8,30)*}$ &$\underline{(6,26)**}$ \\
120& 38 &(2,16) &(20,16)&(20,20) &(22,20) &(26,16) &(22,22) &$\underline{(20,24)}$ &$\underline{(20,22)}$ &(22,16) &$\underline{(14,24)}$ &$\underline{(8,28)}$  
&$\underline{(4,28)*}$ &$\underline{(2,24)**}$ \\
 
\end{tabular}
\end{table*}
\endgroup

For the valence protons of each nucleus, the relevant SU(3) irreducible representations (irreps) of the U(15)$\supset$SU(3) decomposition, obtained by use of the code UNTOU3 \cite{code}, can be seen in Table I (in the hw column). For example, for $^{146}$Ba and $^{168}$Er, which have 6 and 18 valence protons, respectively, the relevant irreps are (18,0) and (18,6), respectively.
  
 In the same table, the relevant SU(3) irreps of the U(21)$\supset$SU(3) decomposition, corresponding to the valence neutrons of each nucleus, can be seen. 
For example, for $^{146}$Ba and $^{168}$Er which have 8 and 18 valence neutrons, respectively, the relevant irreps are (26,4) and (36,6), respectively.  

By taking the sum of these two irreps for each nucleus, 
one can obtain the SU(3) irrep in which the ground state band (and possibly additional bands, according to the value of $\mu$) is located. For example, for $^{146}$Ba and $^{168}$Er one has (18,0)+(26,4)=(44,4) and (18,6)+(36,6)=(54,12), respectively. 

\section{Prolate dominance and the prolate-to-oblate transition} 

The results for the rare earths within the 50-82 proton shell and the 82-126 neutron shell are summarized in Table II. We first note that prolate nuclei are predicted to dominate this region by far, and oblate nuclei only appear at its end in Hf-Pt with large neutron numbers of $N \geq 116$ ($N \geq 118$ for Hf). This prediction is parameter-free in the proxy-SU(3) symmetry and hence constitutes a specific prediction.  

These results are consistent with the experimental data.  First, the overall prolate dominance has been well known for decades, for example, through measurements of quadrupole moments and the success of the Nilsson model on the prolate side. Secondly, while it is more difficult to obtain direct evidence for oblate shapes, quite a number of quadrupole moments are known \cite{Linnemann} in Os-Hg and suggest a shape change at $^{192}$Os$_{116}$-$^{194}$Os$_{118}$. In gamma-soft nuclei, as these are,  oblate shapes develop through a prolate-oblate shape transition passing through a gamma unstable phase, as, for example, in $^{196}$Pt$_{118}$.   Therefore, strong, but indirect, evidence also comes from the systematic behavior of signature observables with neutron number. These are the energy ratio 
$R_{4/2}=E(4^+_1)/E(2^+_1)$ (formation of bubble-like patterns at specific ($N$,$Z$) values), the energy difference $E(2^+_2) - E(4^+_1)$, which crosses zero at the prolate-oblate phase transition, the ratio $E(2^+_2)/E(2^+_1)$, which minimizes for large gamma values, and the $B(E2: 2^+_2\rightarrow 2^+_1)$ value which increases at the shape transition (it is allowed and stronger in gamma-soft nuclei than in well-deformed prolate or oblate nuclei). This shape  transition has been studied in refs. 
\cite{Jolie1,Jolie2,Namenson,Alkhomashi,Wheldon,Podolyak,Linnemann}
 with strong suggestions that  the first oblate nucleus in W is indeed at $^{190}$W ($N = 116$) and in the $^{192-194}$Os isotopes ($N = 116-118$). The requisite data are not yet known for Hf.

The Pt series of isotopes is {\sl not} expected to exhibit the SU(3) symmetry, $^{196}$Pt being the textbook example of the O(6) symmetry
\cite{Cizewski,IA}. However, if one blindly ascribes SU(3) irreps to the series of Pt isotopes, the first oblate one appears to be $^{194}$Pt$_{116}$, 
approximately in  agreement with empirical observations \cite{Jolie1,Jolie2,Linnemann} and theoretical predictions \cite{Sarriguren,Robledo,Nomura83,Nomura84}.
This is also in rough agreement with the empirical observations and theoretical findings of Ref. \cite{Zhang85}, carried out within the SU(3) 
limit of the interacting boson model, locating $^{192}$Pt$_{114}$ near the prolate-oblate transition point. It is also consistent with the expectation that the O(6) symmetry represents the critical point of a prolate-to-oblate shape phase transition \cite{Jolie1,Jolie2}. 

\begingroup
\begin{table*}[htb]

\caption{Same as Table II, but for the most leading SU(3) irreps \cite{Elliott1,Elliott2} for nuclei with protons in the 50-82 shell 
and neutrons in the 50-82 shell. 
}

\bigskip

\begin{tabular}{ r r r | r r r r r r r r r r r r  }

  &  &              & Ba    &  Ce    &  Nd    &  Sm    &   Gd   &  Dy    &  Er    &  Yb    &  Hf    &   W    &   Os   &  Pt     \\
  &  &  $Z$         &  56   & 58     & 60     & 62     & 64     & 66     & 68     & 70     & 72     & 74     & 76     & 78      \\
  &  &$Z_{val}$     &   6   &  8     &  10    &  12    &  14    &  16    &  18    &  20    & 22     &  24    &   26   &   28    \\
$N$ &$N_{val}$& irrep        &(18,0) & (18,4) &  (20,4)& (24,0) & (20,6) & (18,8) & (18,6) & (20,0) & (12,8) & (6,12) & (2,12) & (0,8)   \\   

\hline

56 & 6  &(18,0) &(36,0) &(36,4) &(38,4) & (42,0) &(38,6) & (36,8) & (36,6) & (38,0)& (30,8) & (24,12) & (20,12) & (18,8) \\
58 & 8  &(18,4) &(36,4) &(36,8) &(38,8) &(42,4) &(38,10) &(36,12) &(36,10) &(38,4) &(30,12) &(24,16) &(20,16) &(18,12)  \\
60 & 10 &(20,4) &(28,4) &(38,8) &(40,8) &(44,4) &(40,10) &(38,12) &(38,10) &(40,4)&(32,12) &(26,16) &(22,16) &(20,12)  \\
62 & 12 &(24,0) &\bf{(42,0)}&(42,4) &(44,4) &(48,0) &(44,6) &(42,8) &(42,6) &(44,0) &(36,8) & (30,12) &(26,12) & (24,8) \\
64 & 14 &(20,6) &\bf{(38,6)}&\bf{(38,10)}&(40,10)&(44,6) &(40,12)&(38,14)&(38,12)&(40,6)&(32,14)&(26,18)&(22,18)& (20,14)\\
66 & 16 &(18,8) &\bf{(36,8)}&\bf{(36,12)}&(38,12) &(32,8)&(38,14)&(36,16)&(36,14)&(38,8)&(30,16)&(24,20)&(20,20)& (18,16)  \\
68 & 18 &(18,6) &\bf{(36,6)}&\bf{(36,10)}&\bf{(38,10)}&(42,6)&(38,12)&(36,14)&(36,12)&(38,6)&(30,14)&(24,18)&(20,18)&(18,14) \\
70 & 20 &(20,0) &(38,0)*&\bf{(38,4)}&\bf{(40,4)}&\bf{(44,0)}&(40,6)&(38,8)&(38,6)&(40,0)&(32,8)&(26,12)&(22,12) &(20,8)  \\
72 & 22 &(12,8) &(30,8)* &(30,12)* &\bf{(32,12)}&\bf{(36,8)}&(32,14) &(30,16)&(30,14)&(32,8)&(24,16)&$\underline{(18,20)}$&$\underline{(14,20)}$ &$\underline{(12,16)}$ \\
74 & 24 &(6,12)&(24,12)*&(24,16)* &(26,16)* &(30,12)* &(26,18)* &\bf{(24,20)} &(24,18)&(26,12)&$\underline{(18,20)}$&$\underline{(12,24)}$&$\underline{(8,24)}$&$\underline{(6,20)}$ \\
76 & 26 &(2,12)&        &(20,16)*&(22,16)*&(26,12)* &(22,18)* &(20,20)* &(20,18) &(22,12)&$\underline{(14,20)}$&$\underline{(8,24)}$&$\underline{(4,24)}$&$\underline{(2,20)}$ \\
78 & 28 &(0,8) &        &        &        &         &          &        &(18,14) &(20,8)  &$\underline{(12,16)}$&$\underline{(6,20)}$&$\underline{(2,20)}$& $\underline{(0,16)}$ \\

\end{tabular}
\end{table*}
\endgroup


\begin{figure}[htb]

\includegraphics[width=75mm]{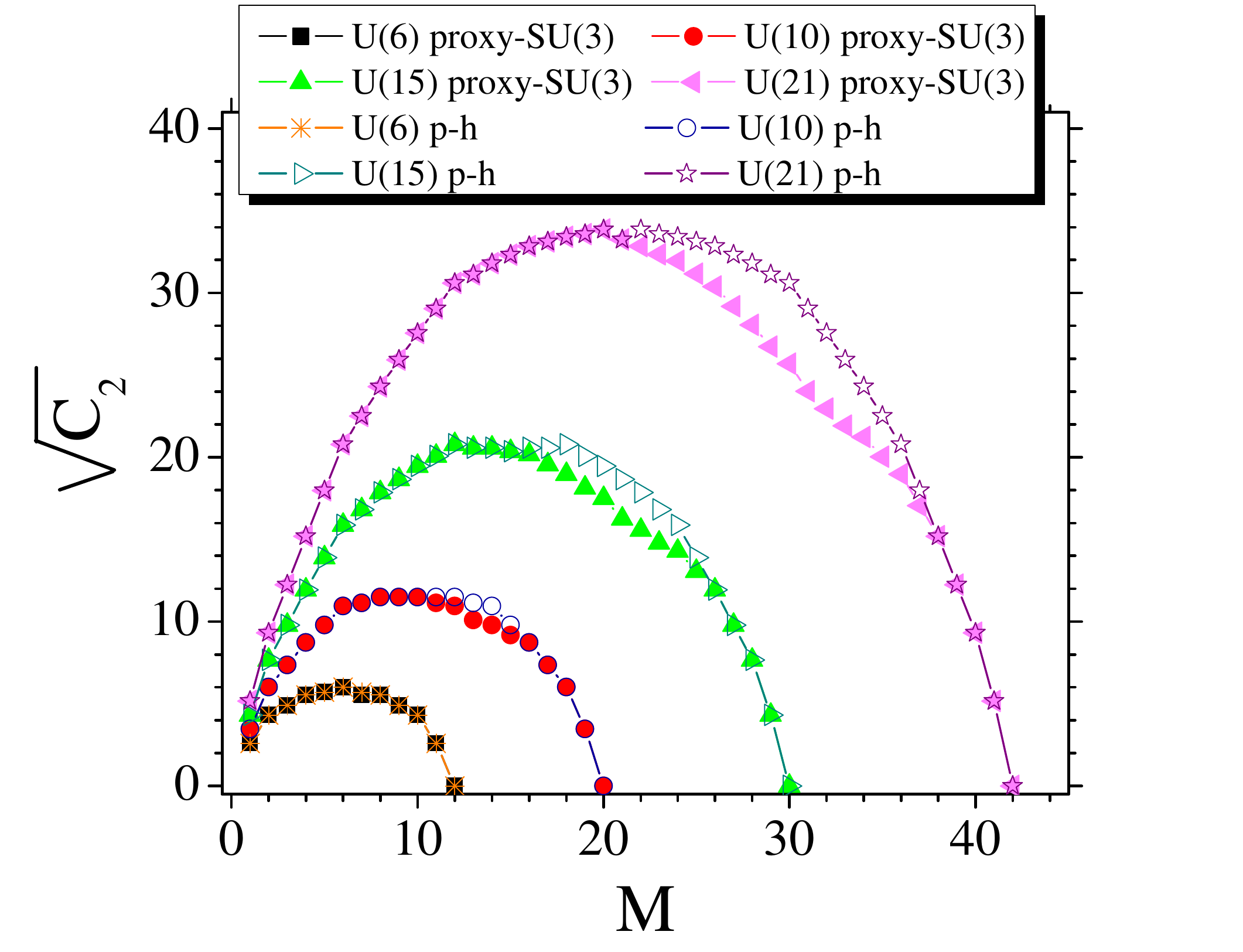}

\caption{Values of the square root of the second order Casimir operator of SU(3), obtained from 
Eq.~(2), vs. particle number M, for different shells, obtained through proxy-SU(3) or through the particle-hole symmetry assumption. See Section \ref{p-h}  for further discussion.} 
\end{figure}

\begin{table}

\caption{Below each shell of the shell model, its proxy-SU(3) content is shown, followed 
by the relevant unitary algebra and the size $S$ of the shell. Further down the highest weight SU(3) irreps are given for nucleon number M. The code UNTOU3 \cite{code} has been used for producing these results.
The sd shell is also shown, up to mid-shell, for comparison. 
See Section \ref{betasc}  for further discussion.}

\bigskip

\begin{tabular}{ r r r r r r r  }

\hline

\hline
   & 8-20 & 28-50 & 50-82 & 82-126 & 126-184 &184-258\\
   & sd   & pf    & sdg   &  pfh   & sdgi    & pfhj  \\
   & U(6) & U(10) & U(15) & U(21)  & U(28)   & U(36) \\
S  & 12   & 22    & 32    & 44     & 58      & 74    \\
M  &      &       &       &        &         &       \\
2 & (4,0) & (6,0) & (8,0) & (10,0) & (12,0) & (14,0) \\     
4 & (4,2) & (8,2) &(12,2) & (16,2) &(20,2)  & (24,2) \\ 
6 & (6,0) & (12,0)&(18,0) & (24,0) & (30,0) & (36,0) \\ 
8 &       & (10,4)&(18,4) & (26,4) & (34,4) &(42,4)  \\
10&       & (10,4)&(20,4) & (30,4) & (40,4) &(50,4)  \\
12&       & (12,0)&(24,0) & (36,0) & (48,0) &(60,0)  \\

\hline

\end{tabular}
\end{table} 

\begin{figure*}[htb]

\includegraphics[width=150mm]{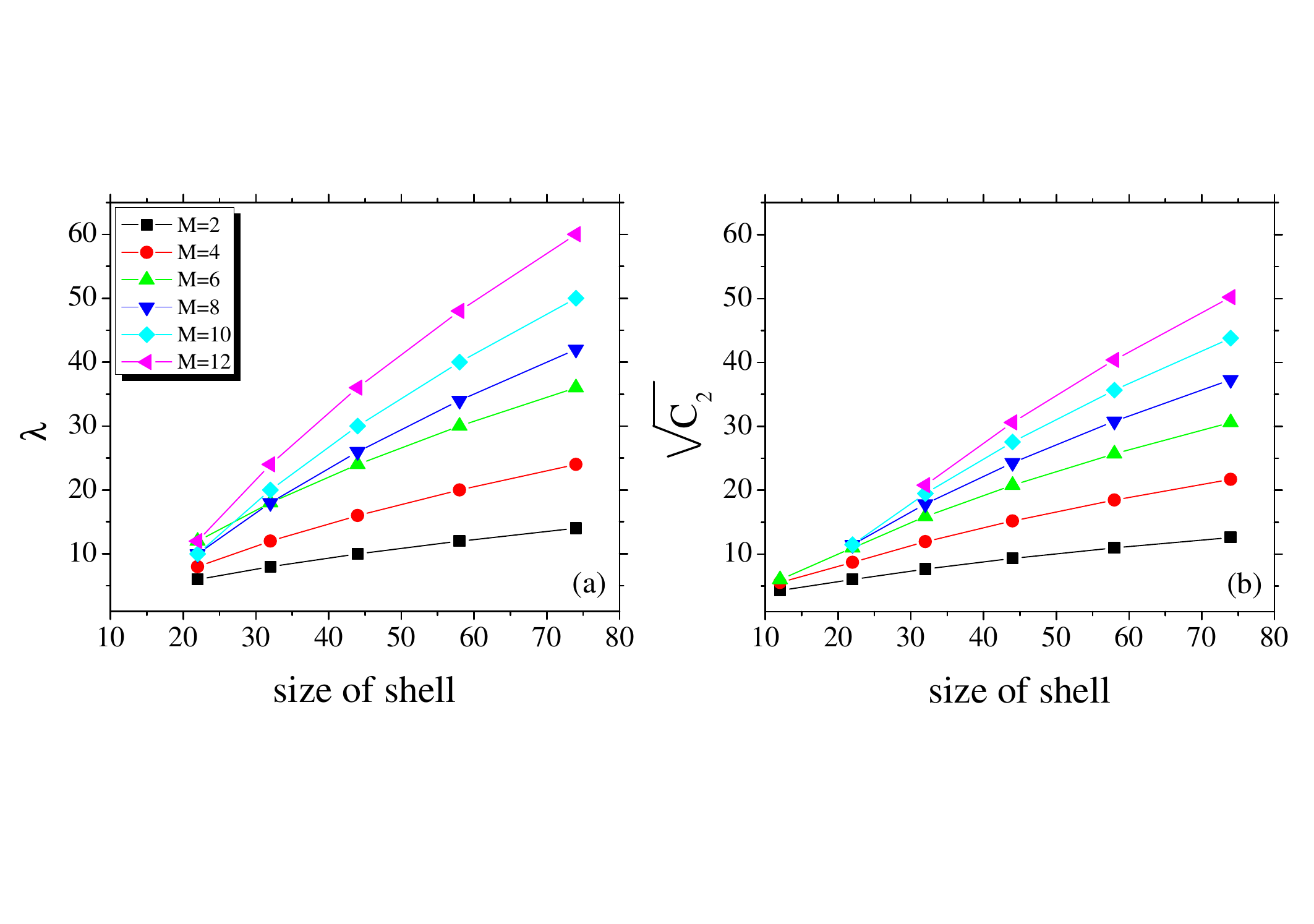}

\caption{(a) Values of the Elliott quantum number $\lambda$ vs. the size of the shell, 
for different nucleon numbers $M$, as shown in Table IV. (b) Values of the square root 
of the second order Casimir operator of SU(3), obtained from Eq.~(2) and related to $\beta$ 
through Eq.~(3), vs. the size of the shell, for different nucleon 
numbers $M$.   See Section \ref{betasc}  for further discussion.} 
\end{figure*}


\begin{figure*}[htb]

\includegraphics[width=150mm]{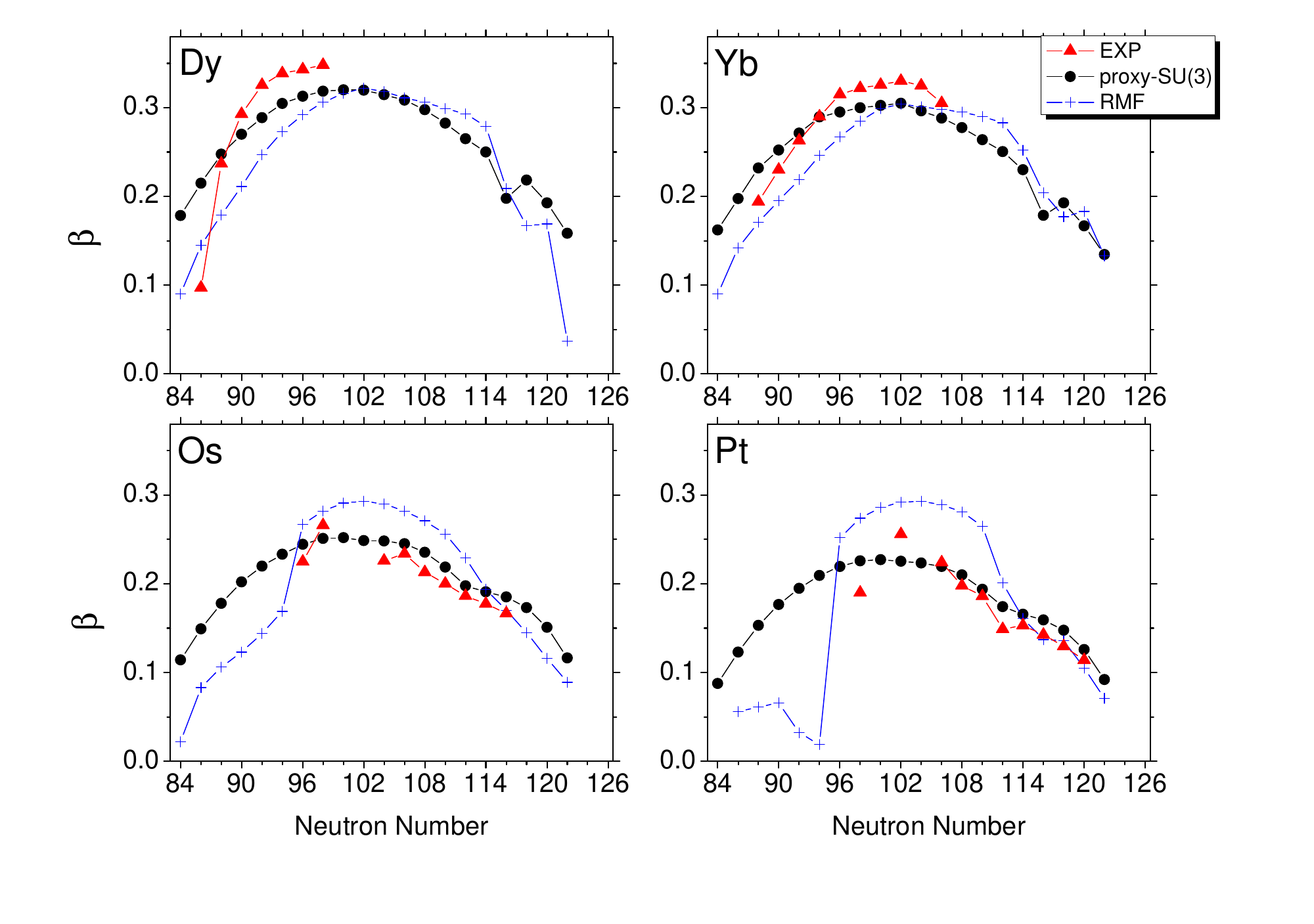}

\caption{Proxy-SU(3) predictions for $\beta$, obtained from Eq.~(3), compared with tabulated $\beta$ values \cite{Raman} and also with predictions from relativistic mean field theory \cite{Lalazissis}. See Section \ref{betasc} for further discussion.
 } 
\end{figure*}


\begin{figure*}[htb]

\includegraphics[width=120mm]{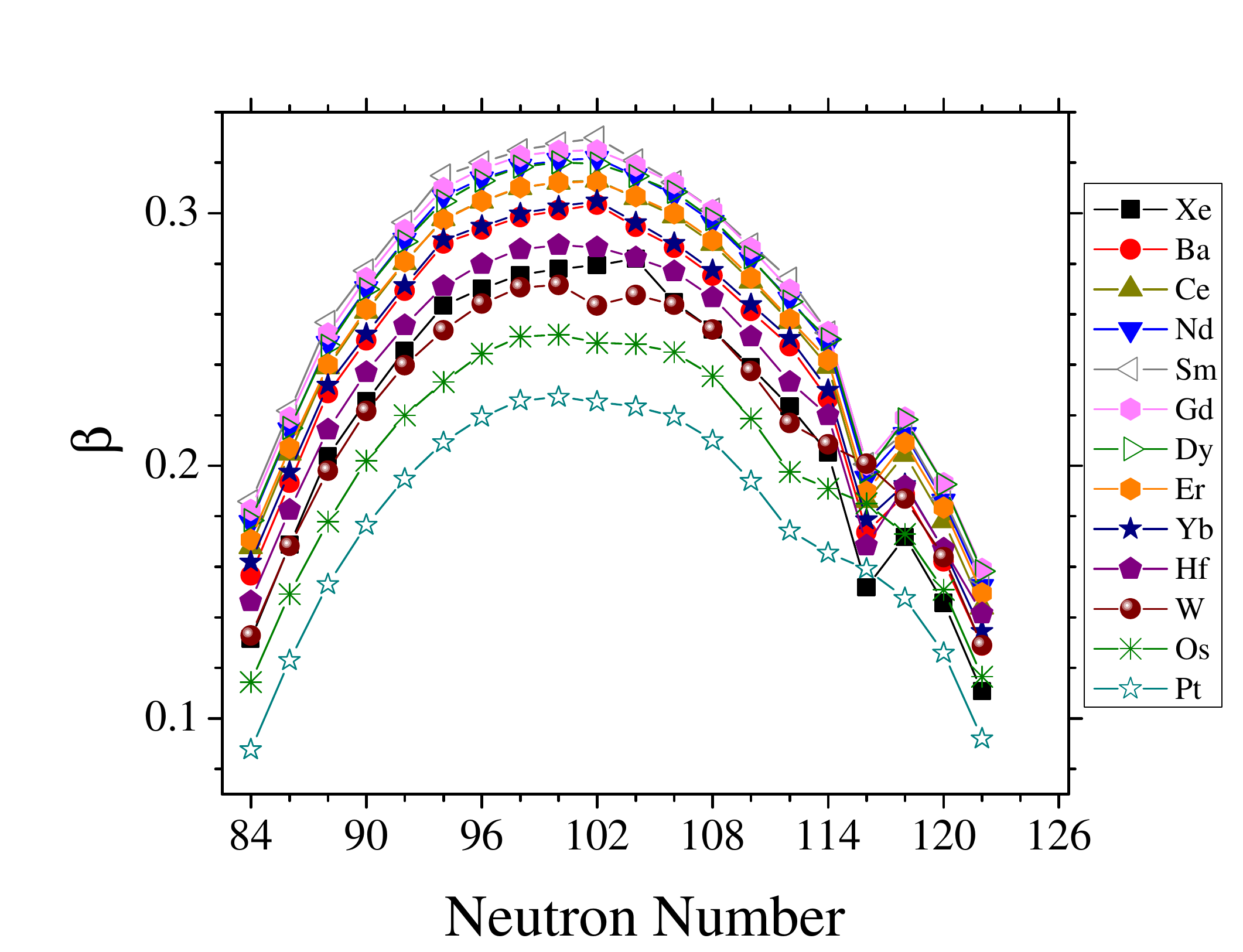}

\caption{Proxy-SU(3) predictions for $\beta$. See Section \ref{betasc} for further discussion.} 

\end{figure*}


\begin{figure*}[htb]

\includegraphics[width=150mm]{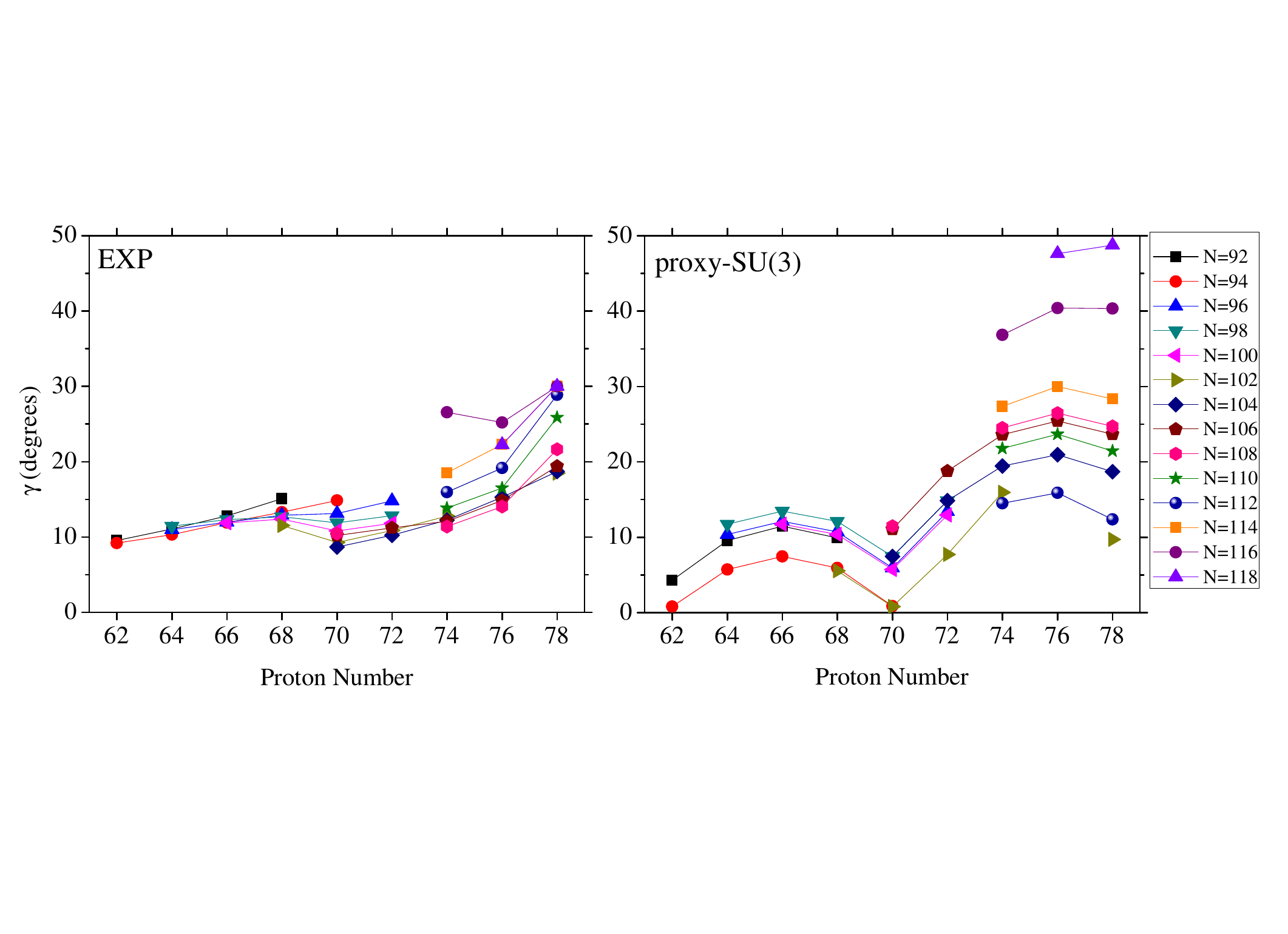}

\caption{Proxy-SU(3) predictions for $\gamma$, obtained from Eq.~(1), compared to empirical values extracted from ratios of the 
$\gamma$ vibrational bandhead to the first $2^+$ state. See Section \ref{Sec-g} for further discussion.
 } 
\end{figure*}


\begin{figure*}[htb]

\includegraphics[width=150mm]{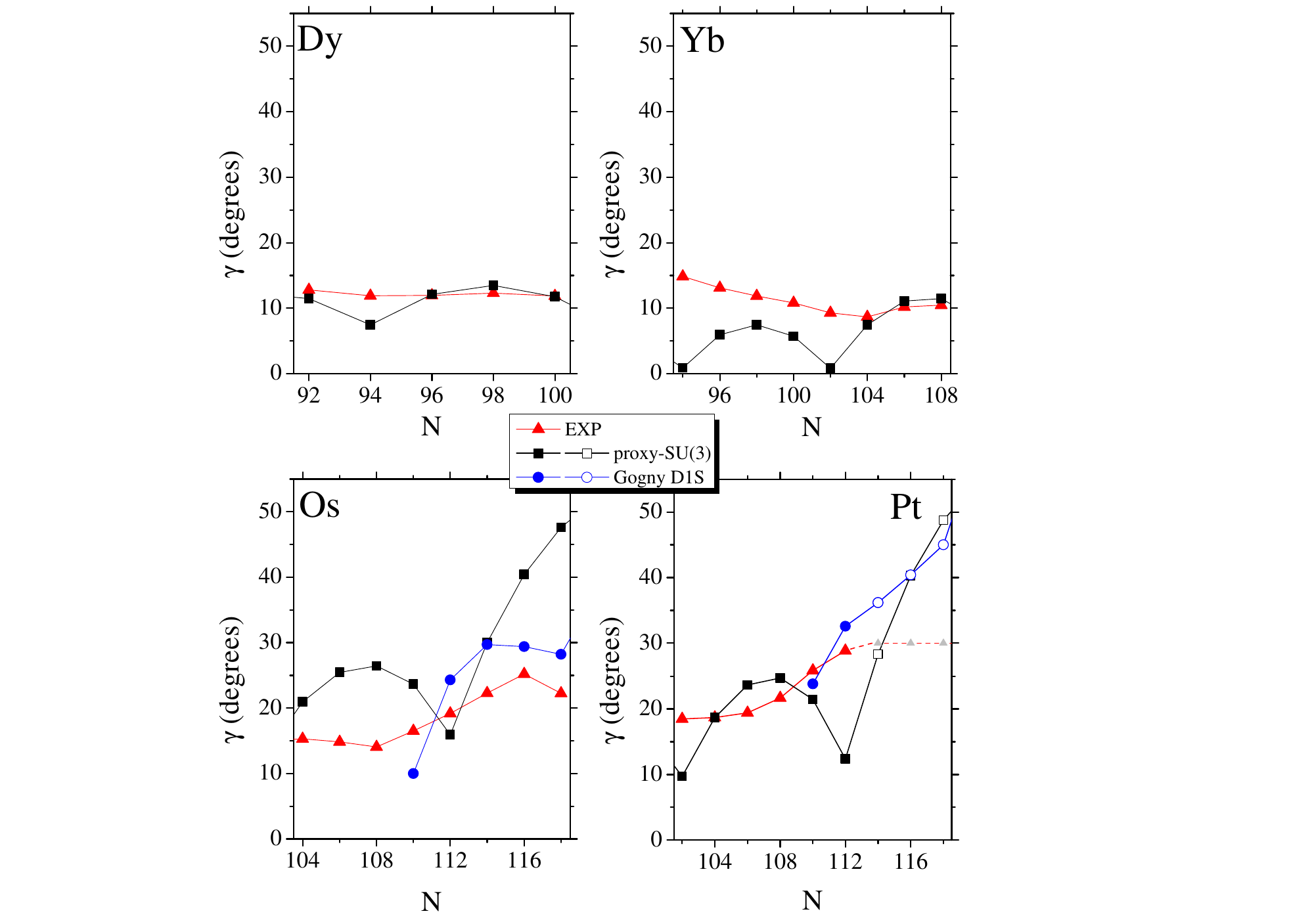}

\caption{Proxy-SU(3) predictions for $\gamma$, obtained from Eq.~(1), compared with 
empirical values obtained from Eq.~(6) \cite{Casten,Esser}, as well as with  predictions
of Gogny D1S calculations \cite{Robledo}. 
The small light triangles for the heaviest Pt isotopes at 30 degrees are used because their experimental values for the ratio $R$ of Eq.~(5) are a few percent below the limiting value for 30 degrees and hence $\gamma$ cannot be rigorously extracted; however there is abundant evidence that these isotopes have asymmetry values near 30 degrees.  The corresponding theoretical values are denoted by open circles.
See Section \ref{Sec-g} for further discussion.}

\end{figure*}

Using the same method one can also consider the rare earths with protons in the 50-82 shell and neutrons also in the 50-82 shell, the results being summarized in Table III. We see that once again the quadrupole deformation maximizes near mid-shell, while a prolate-to-oblate transition also appears at the lower right part of the table. In the W, Os, and Pt series of isotopes, the first oblate nuclei appear at $N = 72$, i.e. they are $^{146}$W, $^{148}$Os, $^{150}$Pt, 
all of them lying far away from the region experimentally accessible at present \cite{ENSDF}, while 
in the Hf series of isotopes, the first oblate nucleus is $^{146}$Hf, having $N=74$. 

These predictions should be considered with extreme care, since in this shell protons and neutrons occupy the same major shell, thus the role of formation of proton-neutron pairs by protons and neutrons occupying identical or very similar orbitals should be examined before any conclusions could be drawn. 

It should be noticed that the prolate over oblate dominance in heavy $N=Z$ nuclei has been recently obtained in the framework of the 
quasi-SU(3) symmetry \cite{Zuker1,Zuker2}, focused in the region from $^{56}_{28}$Ni$_{28}$ to $^{96}_{48}$Cd$_{48}$ \cite{Zuker2}. 

In any case, the generic behavior of the deformation variables is robust and the prolate over oblate dominance is clear in both tables, since in both cases the oblate nuclei are limited to 
the lower right part of the tables, i.e. just below the filling of the proton shell and the simultaneous filling of the neutron shell.

\section{Breaking of particle-hole symmetry} \label{p-h}

The results reported in Table II exhibit clearly that no particle-hole symmetry appears within the framework of proxy-SU(3) symmetry \cite{HM}. From the mathematical point of view, this fact is already made clear by Table I. 
For example, 6 valence protons in U(15) correspond to the (18,0) irrep, while 6 valence proton holes correspond to $82-6=76$ protons, i.e., to 26 valence protons and 
the (2,12) irrep. Similarly, 10 valence neutrons in U(21) correspond to the (30,4) irrep, while 10 valence neutron holes correspond to 
$126-10=116$ neutrons, i.e., to 34 valence neutron holes and the (12,16) irrep. As a result, in Table II one can see that $^{148}$Ba, possessing 6 valence protons and 10 valence neutrons corresponds to the (18,0)+(30,4)=(48,4) prolate irrep, while its particle-hole conjugate, $^{192}$Os, 
possessing 6 valence proton holes and 10 valence neutron holes corresponds to the completely different oblate irrep (2,12)+(12,16)=(14,28). 

We now discuss the particle-hole symmetry breaking in more detail, since it plays a 
crucial role in obtaining the prolate-oblate transition at the right place. 

There are two paths for the selection of the irrep which will be used for the description
of the lowest lying band(s) in a nucleus:

1) In several cases \cite{DW1,DW2}, the irreps are ranked according to the eigenvalue 
of the second order Casimir operator, $C_2(\lambda,\mu)$ of SU(3). The irrep possessing the highest eigenvalue 
of $C_2(\lambda,\mu)$ is supposed to lie lowest in energy, called the most leading irrep. 

2) In proxy-SU(3) \cite{proxy}, the irreps are ranked according to highest weight, as obtained from \cite{code}. Interestingly, this is in accordance with the choice made in the original pseudo-SU(3) work \cite{pseudo1}.

The outcome of the two paths is shown in Table I for several shells.
Up to the middle of the shell, the two paths lead to identical results. 
In the first half of the shell (upper half of the table), the use of the Casimir operator leads to results which are a mirror 
image of the second half of the shell (the lower half of the table), i.e., perfect particle-hole,
and therefore prolate-oblate,  symmetry about mid-shell appears.  

In contrast, the use of the highest weight leads in the first half of the shell (upper half of the table) to results different from the mirror image of the second half of the shell (the lower half of the table), except for the last five irreps in each shell, for which the particle-hole symmetry is valid. As a result, the irreps just below the middle of the table (bold in Table I) are not symmetric [reversed $(\lambda, \mu)$] from those just above the middle of the table. Because of this, the particle-hole symmetry in proxy-SU(3) is destroyed, except for the last 5 irreps. Therefore, the larger the shell, the larger the percentage of irreps breaking the particle-hole symmetry (i.e., the irreps shown in boldface in Table I).

In the work of Elliott \cite{Elliott1,Elliott2,Elliott3} in the sd shell the difference between 
the two paths is almost invisible, since only one irrep is affected, as seen in Table I.
The choice of path also makes no difference up to mid-shell, thus the results obtained 
in the pseudo-SU(3) framework \cite{DW1,DW2} are completely valid.
  
However, the breaking of the particle-hole symmetry in the proxy-SU(3) scheme is instrumental in obtaining the right position 
in the nuclear chart for the prolate-to-oblate transition in the rare earths. 
Without this breaking, the prolate-oblate transition would have taken place in the middle of the shell. 

From the physics point of view, looking at the Nilsson diagrams \cite{Nilsson1,Nilsson2} immediately reveals that particle-hole symmetry is not present. For example, in the 50-82 proton shell, at $\epsilon=0.3$,
the first 8 particles will occupy the  ${1\over 2}$[431],  ${1\over 2}$[550], ${3\over 2}$[422], and ${1\over 2}$[420] orbitals, while the last 8 particles will occupy the ${1\over 2}$[400], 
${3\over 2}$[402], ${11\over 2}$[505], and ${9\over 2}$[514] orbitals. In the last two orbitals 
of the latter case high values of $K$ occur, which do not appear in the former case.
But high $K$ values require high $\mu$ values  in the $(\lambda,\mu)$ irreps of SU(3), in order
to be accommodated, since $K={\rm min}\{ \lambda, \mu\}$, ${\rm min}\{ \lambda, \mu\}-2$, 
${\rm min}\{ \lambda, \mu\}-4$, 1 or 0 \cite{Elliott1,IA}. 

The effect of the particle-hole symmetry breaking is visible in Fig.~1, where the square 
root of the second order Casimir operator of SU(3), which, according to Eq.~(2) is proportional to $\beta$, is plotted as a function of the particle number $M$. Proxy-SU(3) and particle-hole symmetry 
provide identical results up to midshell and at the end of the shell, while between midshell and end
of the shell the differences show up. 

\section{Results for the $\beta$ variable} \label{betasc}

One can go much further in using the proxy-SU(3) than simply the prolate-oblate systematics.
It is also straightforward to use Eqs. (1) and (3) to obtain predictions for the shape variables $\beta$ and $\gamma$ themselves for any given deformed nucleus [any given ($\lambda$, $\mu$)]. These predictions are parameter free (except for the global constant $r_0$ in Eq. (3), for which the value of 0.87 provided in the literature \cite{Castanos,DeVries,Stone} will be used for all nuclei throughout).  Of course, since they are based solely on the SU(3) highest weight irreps, and neglect all interactions except quadrupole, and utilize only the valence shell, they cannot be expected to be very precise. Nevertheless, it is interesting to see what emerges and how well they do work.  

The predictions of the proxy-SU(3) model for the quadrupole deformation $\beta$ will be compared to the detailed predictions of the Relativistic Mean Field theory \cite{Lalazissis}, as well as to experimental values obtained from the $B(E2)$ transition rates from the ground to the first excited $2^+$ state of even-even nuclei \cite{Raman}. 

At this point the question of scaling according to the size of the shell arises. For example, 
in the case of the geometrical limit \cite{GK} of the interacting boson model \cite{IA}, a rescaling factor of $2N_B/A$ is used, where $N_B$ is the number of bosons (half of the number of valence nucleons) present in a nucleus with mass number $A$. In other words, the rescaling factor is 
the number of valence nucleons over the total number of nucleons. 

By analogy here one should have a rescaling factor related to the size of the shells used by the valence protons and valence neutrons as compared to the size of the whole nucleus. From the contents of Table IV and Fig.~2(a) it is clear that (at least for the low values of $M$ shown) $\lambda$ is nearly proportional to the size of the shell, while from Eq. (3) it is clear that (at least in the case of $\lambda \gg\mu$)  $\beta$ is roughly proportional to $\lambda$. As a result, $\beta$ turns out to be roughly  proportional to the size of the shell. If we could accommodate all $A$ particles within a large shell possessing an SU(3) subalgebra, the $\lambda$ of the irrep representing the nucleus would have been proportional to $A$. Here we use the valence protons, for which the relevant 
$\lambda_p$ is proportional to the size $S_p$ of the proton shell, and the valence neutrons, for which the relevant $\lambda_n$ is proportional to the size $S_n$ of the neutron shell. The total irrep 
characterizing the nucleus has $\lambda=\lambda_p+\lambda_n$, therefore $\lambda$ is proportional 
to $S_p+S_n$. This implies that the $\beta$ values obtained from Eq. (3) should be multiplied 
by $A/(S_p+S_n)$. In the case of the rare earth region, where the neutrons fill the 82-126 shell and the protons the 50-82 shell, this gives a re-scaling factor of $A/76$.

Figure~3 shows typical results for $\beta$, for four elements spanning the region, and compares these with tabulated $\beta$ values \cite{Raman} and also with predictions from relativistic mean field theory \cite{Lalazissis}.  Overall, the agreement, both qualitatively and even quantitatively, is surprisingly good given the simplicity of the approach.   For each element, $\beta$ rises to a maximum near mid-shell and then drops off sharply, heading, in the case of Os, towards the prolate-oblate transition discussed above.

In Fig.~4, all the  proxy-SU(3) predictions for $\beta$ for nuclei in the rare earth region are collected. Interestingly, there is a dip in the elements up to Hf at $N=116$, which, for heavier elements, is the locus of the prolate-oblate transition. Moreover, there even appears to be support for the particle-hole symmetry breaking inherent in proxy-SU(3) just after mid-shell. 

\section{Results for the $\gamma$ variable}\label{Sec-g}

Figure~5 shows for this entire region numerical results for the proxy-SU(3) predictions for $\gamma$, compared to empirical values extracted from ratios of the $\gamma$ vibrational bandhead to the first $2^+$ state,  
\begin{equation}
R={E(2^+_2)\over E(2^+_1)}, 
\end{equation}
according to \cite{DF,Casten,Esser}
\begin{equation}
\sin 3\gamma= {3\over 2\sqrt{2}} \sqrt{1-\left({R-1\over R+1}  \right)^2}. 
\end{equation} 
Values extracted from other observables such as $B(E2)$ values can differ by 2-3 degrees giving a feeling for the experimental uncertainties (see Refs. \cite{Casten,Warner}). 

The comparisons are particularly interesting. First, the data in Fig. 5 show two distinct patterns - roughly constant values near 10-15 degrees for well deformed nuclei, and a sharp rise toward maximum axial asymmetry in the Os ($Z=76$) and Pt ($Z=78$) isotopes. 
It is important to note that physical differences in the ground and $\gamma$ band wave functions for $\gamma$ values below about 15 degrees are extremely small. For example, probabilities for $K = 2$ components in the ground state band $4^+$ state are about 1\% or less \cite{DF,Casten,Warner}. 
Hence, differences of a few degrees in predicted and empirical $\gamma$ values in this range are not physically very significant and the qualitative agreement is quite good.

We note that there are, however, two obvious areas of disagreement.  
Near the end of the shell the empirical $\gamma$ values (determined in the way described above) saturate at about 30 degrees (maximum axial asymmetry), while the proxy-SU(3) predictions show a return to axial symmetry for oblate shapes ($\gamma$ approaching 60 degrees). 

Also the empirical values from $Z = 62$ to 72 are rather smooth and gradually increasing, while the predictions show rather strong oscillations.  It is our speculation that this reflects the effects of pairing which are ignored here and which will tend to spread the orbit occupancies in the ground state (and hence in the $\gamma$ mode which is related to the ground state through a $Y_{22}$ operator)  and mute these oscillations. Further work is needed to test this hypothesis.  Nevertheless, despite this discrepancy we note an interesting common feature:  The empirical values do show a drop going to $Z=62$ 
and a soft bottoming out at $Z = 70$, just where proxy-SU(3) also has (more distinct) minima.

In Fig.~6 we show detailed comparisons for four elements spaced out over the region, including, for Os and Pt, the results of Gogny D1S predictions \cite{Robledo}. Overall the agreement is reasonable. Both empirical values and the proxy-SU(3) predictions are uniformly below about 15 degrees for Dy and Yb, although the proxy-SU(3) results show strong oscillations for Yb with a distinct minimum near 
$N = 102$. This latter may be related to the onset of a bosonic SU(3) axial symmetry, which has been argued for in these nuclei near $N = 102$, 104 \cite{Haque}. In Os and Pt, both the empirical values and the predictions agree on a sharp increase in $\gamma$ beyond $N \sim 110$. The proxy-SU(3) results are certainly no worse than those using a Gogny D1S interaction \cite{Robledo}, but the agreement is only qualitative. 

The same analysis can be carried out for the $Z = 50$-82, $N = 50$-82 shell. Again, the axial asymmetry variable $\gamma$ has modest values around 10-15 degrees near mid-shell, and rises towards 30 degrees near the end of the shell. We thus see that the behavior of these variables is similar in different shells, as again reflected in abundant data.

Finally, we again stress the extreme simplicity of these predictions, based on an approximate group structure and on the filling of nucleons in a quadrupole field. No account is taken of pairing, the contributions of other shells, or other higher multipole or other interactions. In particular, one expects the effects of pairing to wash out the predicted oscillations as pairing scatters pairs of particulars among several orbitals near the Fermi surface.  Taking pairing into account would presumably therefore improve the agreement. We are pursuing an implementation of pairing but that is beyond the scope of the current paper.

\section{Conclusions}

In the present work the prolate over oblate dominance in deformed rare earth nuclei is obtained within the framework of a parameter-free proxy 
SU(3) symmetry, using the symmetry properties alone. In addition, within the same SU(3) framework, 
the point of the prolate-oblate shape phase transition is predicted to be at $N \approx 116$, 
as supported by the existing experimental data  and recent microscopic calculations. 
Finally, we have used Eqs. (1) and (3) to obtain simple, analytic, parameter-free (except for 
a single global value of $r_0$ in the case of $\beta$) predictions of the $\beta$ and $\gamma$ deformation variables for the rare earth region.  Similar results for other regions are trivially obtainable once the irreps (analogous to Table II) are obtained.  The predictions are broadly consistent with the empirical results. The quadrupole deformation $\beta$ has the observed roughly parabolic behavior across deformed nuclei. For $\gamma$, the empirical results of values near 10-15 degrees for most of the deformed rare earth nuclei and the sharp rise towards 30 degrees at the upper end is also reproduced.  Clearly the predictions show oscillations not evident, or at least highly muted, in the data. We suspect that this is due to the neglect of pairing for which pair scattering would have the tendency to smooth out  the variations, but that needs to be checked by further study.

SU(3) symmetry in nuclei is known to be connected to the dominance of the quadrupole-quadrupole interaction 
\cite{Elliott1,Elliott2,DW1,DW2}. Therefore one could think that the prolate over oblate dominance in the deformed rare earths,
as well as the prolate-oblate shape phase transition are direct consequences of the quadrupole-quadrupole interaction dominance. 

It should be remembered that the Nilsson levels are {\sl not} changed much by the approximation involved
in the proxy-SU(3) scheme \cite{proxy}, since in each shell the normal parity orbitals remain intact, while the intruder parity orbitals are replaced by their 0[110] partners. In particular, downwards leading prolate orbits remain downwards leading, while upwards moving oblate orbits remain upwards moving. Moreover, the enhanced mixing of single particle states on the prolate side also persists, leading to avoided crossings and a further lowering of the prolate states \cite{Casten}.
As a result, it is expected that this simple approximate scheme should provide results 
consistent with microscopic treatments concerning nuclear properties related to the prolate or oblate character of deformed nuclei.   

The present work suggests that it is worth investigating how far one can go in the description of the properties of heavy deformed nuclei taking advantage of the proxy-SU(3) symmetry scheme.
There are, however, important caveats about what this approach can and cannot accomplish and about its inherent limitations. It invokes, through SU(3), the valence space quadrupole interaction.  Hence effects such as pairing, or a larger space involving additional major oscillator shells, or more complex interactions, are so far ignored. In no way can it replace large scale shell model calculations, ab initio, first principles, methods, nor is it intended to do so. It is complementary to such microscopic approaches and is adept at predicting, in a very simple way, as illustrated in this paper,  those observables that robustly depend primarily on valence nucleon number and the quadrupole interactions amongst them. In this first application of the proxy-SU(3) scheme, we feel that  the simplest approach is reasonable, in order to see what the model can do on its own. To go further, of course, some ways of incorporating at least some additional degrees of freedom, such as pairing, will need to be studied. For now, we feel that the success of the proxy-SU(3) has exhibited its potential for the prediction of global nuclear properties and that this should encourage further study of it and its possible extensions.

A possible path is briefly mentioned here. 
For the nuclei shown in Table II the most leading SU(3) irrep will contain bands with 
$K=\mu$, $\mu-2$, \dots, 0 \cite{Elliott2}. As a consequence, in principle several bands will coexist within the most leading irrep. In the case of $\mu=6$, for example, the $K=0$ ground state band, the $K=2$ $\gamma_1$ band, the first $K=4$ band and the first $K=6$ band will coexist. The degeneracy among these bands can be broken through the use of three-body and/or four-body terms which are known to be O(3) scalars belonging to the SU(3)$\supset$O(3) integrity basis \cite{DW1,DW2,PVI}. Although this goes beyond proxy-SU(3) per se, since it involves additional interactions, the proxy-SU(3) starting point may simplify such calculations.  Work along these lines is in progress.  

\section*{Acknowledgements} 

The authors are grateful to J. P. Draayer for providing the code UNTOU3 and for helpful discussions.
Support by the Bulgarian National Science Fund (BNSF) under Contract No. DFNI-E02/6 is gratefully acknowledged by N. M.
Work supported in part by the US DOE under Grant No. DE-FG02- 91ER-40609, and by the MSU-FRIB laboratory. R.B.C. acknowledges support from the Max Planck Partner group, TUBA-GEBIP, and Istanbul University Scientific Research Project No. 54135.

\section*{Appendix A: The origins of Eqs.~(1) and (3)}

In this Appendix the concepts behind the derivation of Eqs. (1) and (3) \cite{Evans,Castanos,Park} are discussed.

Eqs.~(1) and (3) are obtained through a linear mapping \cite{Castanos,Park} between the eigenvalues of the invariant operators of the 
collective model \cite{BM}, $\beta^2$ and $\beta^3\cos 3\gamma$, and the eigenvalues of the invariants of SU(3). The basic idea behind this approach is that if the invariant 
quantities of two theories are used to describe the same physical phenomena, their measures  
must agree \cite{Evans,Castanos,Park}. The derivation is based on the fact that SU(3) contracts 
\cite{Rowe} to the quantum rotor algebra \cite{Ui}
for low values of angular momentum and large values of the second order Casimir operator of SU(3),
the eigenvalues of which are given in Eq. (2). 
Large values of $C_2$ occur for large values of $\lambda$ and/or $\mu$. The SU(3) irreps
appearing in Tables II and III do have $\lambda$ and/or $\mu$ large, thus the use of the 
contraction limit is justified in these cases. Furthermore, it turns out that to each 
$(\lambda,\mu)$ irrep corresponds a unique value of the $(\beta, \gamma)$ variables. 

\section*{Appendix B: The U(N)$\supset$SU(3) decompositions}

In this Appendix we briefly explain the way in which the results reported in Table I are obtained. 

\subsection*{1. The U(15)$\supset$SU(3) decomposition}

In the U(15) algebra the fundamental irreducible representation (irrep) is [1] (a Young tableau with one box). Since U(15) is the algebra describing the sdg shell, the corresponding SU(3) irrep should 
contain the angular momenta $L=0$, 2, 4. It is known \cite{Elliott1} that the angular momentum 
eigenvalues appearing within an irrep $(\lambda, \mu)$ with $K=0$ are 
\begin{equation}\label{Lvalues}
L=\max\{\lambda,\mu\},\\ \max\{\lambda,\mu\}-2, \\ \ldots, \\ 1 \quad \textrm{or} \quad 0. 
\end{equation}
Therefore in this case the SU(3) irrep has to be (4,0). 

The outer product \cite{Wybourne} of U(15) irreps $[1]\otimes [1]$ results in the  symmetric irrep [2] and the antisymmetric irrep [11]. Using the standard techniques \cite{Wybourne} of calculating the outer product 
of SU(3) irreps $(4,0)\otimes (4,0)$ one finds that the symmetric U(15) irrep [2] contains 
the SU(3) irreps (8,0), (4,2), (0,4), while the antisymmetric U(15) irrep [11] contains the SU(3) irreps
(6,1) and (2,3). Among these SU(3) irreps, the most leading one, 
defined as the one possessing the highest value of the second order Casimir operator of SU(3), 
given in Eq. (\ref{C2}), is (8,0), belonging to the irrep [2] of U(15). 

Since we are considering a system of protons {\sl or} neutrons, larger U(15) irreps will be limited 
by the Pauli principle to a maximum of two columns (allowed because of the two possible orientations 
of spin). Soon enough the calculation for larger irreps becomes cumbersome, thus one has to rely 
on the computational method described in Ref. \cite{code}.  An earlier tabulation of several results 
for the U(15)$ \supset $SU(3) decomposition have been given in Ref. \cite{Brahmam}.  
 
\subsection*{2. The U(21)$\supset$SU(3) decomposition} 
 
The decomposition of U(21) is similar. Since U(21) is the algebra describing the pfh shell, the corresponding SU(3) irrep should contain the angular momenta $L=1$, 3, 5.  Therefore in this case from Eq. (\ref{Lvalues}) one sees that  the SU(3) irrep has to be (5,0). 

The outer product \cite{Wybourne} of U(21) irreps $[1]\otimes [1]$ results in the  symmetric irrep [2] and the antisymmetric irrep [11]. Calculating the outer product 
of SU(3) irreps $(5,0)\otimes (5,0)$ one finds that the symmetric U(21) irrep [2] contains 
the SU(3) irreps (10,0), (6,2), (2,4), while the antisymmetric U(21) irrep [11] contains the SU(3) irreps
(8,1), (4,3), and (0,5). The most leading SU(3) irrep is (10,0), belonging to the irrep [2] of U(21).

\end{document}